\documentclass[aps,prd,twocolumn,showpacs,groupedaddress,nofootinbib]{revtex4}

\usepackage{graphicx}
\usepackage{longtable}

\begin{document}

\title{\bf 
Hamiltonian Study of Improved $U(1)$  Lattice Gauge Theory in Three 
Dimensions}

\author{Mushtaq Loan}
\email{mushe@phys.unsw.edu.au}
\author{Chris Hamer}

\affiliation{School of Physics, The University of New South Wales,
Sydney, NSW 2052, Australia}

\date{\today}
     
\begin{abstract}
A comprehensive analysis of the Symanzik improved anisotropic 
three-dimensional $U(1)$ lattice gauge theory in the Hamiltonian limit is 
made. Monte Carlo techniques are used to obtain numerical results for
the static potential, ratio of the renormalized and bare anisotropies,   
the string tension, lowest glueball masses and the mass ratio.
Evidence that rotational symmetry is established more accurately for 
the Symanzik improved anisotropic action is presented. The
discretization errors in the static potential and the renormalization of
the bare anisotropy are found to be only a few percent compared to  errors
of about 20-25\% for the unimproved gauge action.  
Evidence of scaling in the string tension, antisymmetric 
mass gap and the mass ratio is observed in the weak 
coupling region and the behaviour is tested against  
analytic and numerical results obtained in various other Hamiltonian 
studies of the theory. We find that more accurate determination of the 
scaling  coefficients of the string tension and the antisymmetric mass gap 
has been achieved, and the agreement with various other Hamiltonian studies 
of the theory is excellent.
The improved action is found to give faster 
convergence to the continuum limit. Very clear evidence is obtained that in the
continuum limit the glueball ratio $M_{S}/M_{A}$ approaches exactly 2, 
as expected in a theory of free, massive bosons.
\end{abstract}


\maketitle

\section{Introduction}
Lattice gauge theory calculations have demonstrated important qualitative 
features of QCD, with increasing accuracy. Most lattice gauge theory 
calculations to date have been performed using Monte Carlo techniques in the
Euclidean formulation. Although the Euclidean lattice 
gauge theory has been a very successful non-perturbative technique to compute 
the properties of elementary particles over the years, there are areas where 
progress has been very slow. Examples are  QCD at finite temperature, 
glue thermodynamics, heavy quark spectra etc. Some of these problems have 
resisted  solution even by the powerful techniques of Euclidean field 
theory. This suggests that alternative methods should be pursued in parallel 
with Euclidean lattice gauge theory. A viable alternative that needs to be 
explored  is the Hamiltonian version of QCD. This approach  provides 
a valuable check of the universality of the Euclidean results 
\cite{hamer96} and has an appealing aspect in reducing lattice gauge theory to 
a many-body problem. As such the formalism is
suited for the application of a host of analytic methods imported from quantum 
many body theory and condensed matter physics. It has been suggested that
Hamiltonian lattice gauge theory could more readily handle finite density QCD
\cite{gregory00}. The problems encountered in  finite density QCD in 
the Euclidean formulation have prompted a return to the strong coupling 
expansions of early Hamiltonian lattice gauge theory \cite{bringoltz02}.
Similar ideas have been pursued recently by Luo {\emph et al}.,
\cite{luo02,guo02} who  propose an alternative Hamiltonian lattice 
formulation, `the Monte Carlo Hamiltonian', and have already demonstrated its 
validity and efficiency for the $\Phi^{4}$ model \cite{huang02}.

Here we attempt to extend the standard Euclidean path integral Monte Carlo 
techniques to Symanzik improved $U(1)$ gauge theory in three dimensions.
Applications of this method have been extremely
successful \cite{cre79,cre83} and have given rise to great optimism about 
the possibility of obtaining results relevant to continuum physics from 
Monte Carlo simulations of lattice versions of the corresponding theory. 
Such an optimistic view is supported 
by recent work of Sexton \emph{et al.}  \cite{sexton95}, Boyd \emph{et al.}  
\cite{boyd96}, Luo \emph{et al.}  \cite{xqluo00} and Morningstar and 
Peardon \cite{morn99a} who have attempted to derive masses of the low-lying 
hadrons from such calculations and report successful results. More recent 
is the application of PIMC techniques to obtain  results in 
the Hamiltonian limit for the $U(1)$ model in  (2+1) dimensions \cite{mushe02} 
and SU(3)  lattice gauge theory in (3+1) dimensions on anisotropic 
lattices \cite{tim03}. 

The use of improved actions 
\cite{symanzik83,lusher85} makes possible accurate
Monte Carlo simulations of QCD on coarse lattices with greatly reduced
computational effort \cite{morn99a,morningstar97,morningstar99b}.
In principal, with  an improved action it is possible to achieve lattice 
volumes large enough to overcome finite-size effects and obtain
measurements with good statistical errors.
Coupled with  tadpole improvement \cite{lepage93}, the 
pursuit of the Symanzik program has led to significant progress in reducing
the discretization errors and the renormalization of the anisotropy to the
level of a few percent;  makes using anisotropic lattices no more this
difficult than isotropic ones. At the same 
time the merits of using an improved anisotropic lattice have been well 
understood \cite{morningstar97,sakai00}. Anisotropic lattices allow us
to carry out numerical simulations with a fine temporal resolution while
keeping the spatial lattice spacing coarse, i.e., $a_{t}<a_{s}$, where 
$a_{t}$ and $a_{s}$ are the lattice spacings in the temporal and spatial 
directions, respectively. This is especially important for QCD Monte Carlo 
simulations at finite temperature and heavy particle spectroscopy. But more 
importantly it should make extrapolations to the continuum limit more 
reliable.

As mentioned above, our aim  in this work is to  apply standard Euclidean 
path integral Monte Carlo techniques to extract the Hamiltonian limit for 
Symanzik improved $U(1)_{2+1}$ lattice gauge theory on anisotropic lattices.
The idea is to measure physical quantities on increasingly anisotropic 
lattices, and make an extrapolation to the extreme anisotropic limit.
The effect of the plaquette improvement is examined by
studying the scaling behaviour and the sensitivity of scaling 
coefficients of the string tension and glueball masses in the Hamiltonian 
limit, $\xi\rightarrow 0$. 
The rest of the paper is organized as follows:
In Sec. II we briefly review the formulation of the Symanzik improved 
$U(1)$ gauge action in three dimensions on an anisotropic lattice. The details 
of the simulations and the methods used to
extract the observables are described in Sec. III. Here we discuss our 
techniques for calculating the static quark potential, renormalization 
of anisotropy, string tension and glueball
masses from Wilson loop operators. 
We present and discuss our results in Sec. IV. Scaling of the string 
tension, antisymmetric mass gap and the mass ratio in the weak-coupling 
region, are tested against  theoretical predictions and compared  
with the estimates obtained by other studies in the Hamiltonian limit. 
Our conclusions are given in Sec. V, along with an outline of future work.

\section{Improved Anisotropic Discretization of $U(1)_{2+1}$}

The Symanzik improved $U(1)_{2+1}$ gauge action 
on an anisotropic lattice is identical in form to the SU(3) case 
and is given\footnote{The notation used here differs slightly from 
that used
in Ref. \protect\cite{alford95}, where the prefactors were absorbed
into the definitions of $\beta$ and $\xi$. We follow the notation 
introduced in \protect\cite{morningstar97}.}
by \cite{morningstar97} 
\begin{eqnarray}
S_{g}& = & 
\beta_{s}\xi\sum_{x}\sum_{i<j}\left[\frac{5}{3u^{4}_{s}}P_{ij}(x)
-\frac{1}{12u^{6}_{s}}\big(R_{ij}(x)+R_{ji}(x)\big)\right]
\nonumber\\
& &  +\frac{\beta_{t}}{\xi}\sum_{x}\sum_{i}\left[\frac{4}{3u^{2}u^{2}_{t}}
P_{it}(x)-\frac{1}{12u^{4}_{s}u^{2}_{t}}R_{it}(x)\right],
\label{eqn1}
\end{eqnarray}
where $P_{\mu\nu}$ and $R_{\mu\nu}$ are the $1\times 1$ Wilson loop and 
$2\times 1$ rectangular loop in the $\mu\times\nu$ plane respectively. 
At the tree-level the coefficients are chosen so that the action has no 
$O(a^{2})$ discretization corrections. The spatial and temporal square 
and rectangular loops are given by
\begin{eqnarray}
P_{ij}(x)&=& \left[1-U_{i}(x)U_{j}(x+\hat{i})U^{\dagger}_{i}(x+\hat{j})
U^{\dagger}_{j}(x)\right]\\
R_{ij}(x)&=& \left[1-U_{i}(x)U_{i}(x+\hat{i})U_{j}(x+2\hat{i})
U^{\dagger}_{i}(x+\hat{i}+\hat{j})
\right.
\nonumber\\
& & \left.\times U^{\dagger}_{i}(x+\hat{j})U^{\dagger}_{j}(x)\right]\\
P_{it}(x)&=& \left[1-U_{i}(x)U_{t}(x+\hat{i})U^{\dagger}_{i}(x+\hat{t})
U^{\dagger}_{t}(x)\right]\\
R_{it}(x)&=& \left[1-U_{i}(x)U_{i}(x+\hat{i})U_{t}(x+2\hat{i})
U^{\dagger}_{i}(x+\hat{i}+\hat{t})
\right.
\nonumber\\
& & \left.\times U^{\dagger}_{i}(x+\hat{t})U^{\dagger}_{t}(x)\right],
\label{eqn2}
\end{eqnarray}
where $x$ labels the sites of the lattice, $i$, $j$ are the spatial indices
and $U_{i}(x)$ is the link variable from site $x$ to $x+\hat{i}$.
The rectangular loop that extends two steps in 
the time direction has not been included in the above action. This 
has the advantage of eliminating negative
residue high energy poles in the gluon propagator \cite{morningstar97} but
at the same time leaves errors of the order $a^{2}_{t}$ in the action. 
The $O(a^{2}_{t})$ errors, however,  are negligible provided $a_{t}$ is 
small compared to $a$. The
bare anisotropy parameter is $\xi$ and is equal to the aspect ratio of the 
temporal and spatial lattice spacings at the tree level. At higher orders in 
the perturbative expansion, the bare anisotropy in the simulated action is 
not the same as the measured value, $\xi_{phys}$,
due to quantum corrections.
The couplings $\beta_{s}$ and $\beta_{t}$ are defined by
\begin{equation}
\beta_{s}=\frac{1}{g^{2}_{s}}, \hspace{1.0cm}\beta_{t}=\frac{1}{g^{2}_{t}}
\label{eqn6}
\end{equation}
The two different couplings in Eq. (\ref{eqn1}) are necessary in order to 
ensure that in the continuum limit, physical observables become independent of 
the kind of lattice regularization chosen.
In the case of an asymmetric 
lattice, this implies that physical quantities have to be independent of the 
anisotropy factor $\xi$. To achieve this one needs to introduce 
different couplings for spatial and temporal directions, which depend on $\xi$.
The $\xi$-dependence of  the couplings $g^{2}_{s}$ and $g^{2}_{t}$
is due to quantum corrections and leads to an energy sum rule for the 
quark-antiquark potential, and the glueball mass, which differs in an 
important way from that which one would expect naively.

In the weak coupling limit of $SU(N)$ lattice gauge theory, the relation 
between 
the scales of the couplings in Euclidean and Hamiltonian formulations 
has been determined analytically from the effective actions 
\cite{Karsch82,hasenfratz81,kawai81,hamer95}, using 
the background field method on the lattice. For small  values of 
$g_{E}$, the couplings $g_{s}$ and $g_{t}$ can be expanded as
\begin{eqnarray}
\frac{1}{g^{2}_{t}}& = & \frac{1}{g^{2}_{E}}+c_{t}(\xi )+O(g^{2}_{E})\\
\frac{1}{g^{2}_{s}}& = & \frac{1}{g^{2}_{E}}+c_{s}(\xi )+O(g^{2}_{E})
\label{eqn8}
\end{eqnarray}
where $g_{E}$ is the Euclidean coupling. 
For $\xi =1$, one recovers the usual Euclidean lattice gauge theory, 
where $g_{s}=g_{t}=g_{E}$.
In the limit $\xi \rightarrow 0$, Eqs. (7) and (8) reduce to their 
Hamiltonian values and one obtains the relation between the Euclidean 
coupling $g_{E}$ and its Hamiltonian counterpart $g_{H}$. Similar 
calculations have been performed to determine  the anisotropic coefficients 
$c_{s}$ and $c_{t}$ at arbitrary anisotropy  
for a class of improved actions \cite{margarita97,sakai99,engels00,sakai00}. 
These coefficients become an 
important tool in the analysis of glue thermodynamics \cite{hashimoto93}, the 
quark-gluon plasma \cite{nakamura96,nakamura98} and for the determination of 
spectral functions at finite temperature \cite{forcrand98}. Similar
calculations have not yet been done for the $U(1)$ theory, however, as far as
we are aware.

Tadpole improvement \cite{lepage93} is introduced by renormalizing the 
link variables: $U_{i}(x)\rightarrow U_{i}(x)/u_{s}$, and $U_{t}(x)\rightarrow 
U_{t}(x)/u_{t}$, where the mean fields $u_{s}$ and $u_{t}$ can be defined by 
using the measured values of the average plaquettes in a simulation. In the 
plaquette mean-link formulation, the mean fields are determined 
self-consistently and are defined by
\begin{equation}
u_{s}= (P_{ij})^{1/4}, \hspace{0.50cm} u^{2}_{t}u^{2}_{s}=P_{it}.
\label{eqn9}
\end{equation}
For $a_{t}<<a$, the mean temporal link $u_{t}$ is expected to be very close 
to unity. For simplicity we use a convenient and 
gauge invariant definition for $u_{s}$ in 
terms of the mean spatial plaquette given by 
$u_{s}= \langle \mbox{Re}P_{ij}\rangle^{1/4}$, and compute $u_{t}$ 
from the temporal plaquette $P_{it}$ in Eq. (\ref{eqn9}).

To obtain the Hamiltonian estimates from  anisotropic lattices, a naive
extrapolation procedure is followed. In this procedure we assume 
classical values of the couplings,
i.e., $\beta =\beta_{s}=\beta_{t}$ in Eq. (\ref{eqn1})  and extrapolate 
the  physical quantities  
to the extreme anisotropic limit, $\xi\rightarrow 0$ at constant $\beta$.
Such a procedure is not strictly correct, 
however, at the quantum level because $\beta_{s}\neq\beta_{t}\neq\beta$ due to 
renormalization\footnote{One-loop calculations of the renormalization of
the anisotropy and the gauge  coupling in spatial and temporal directions for 
the improved Abelian lattice gauge theory are currently under way and will be 
reported elsewhere \cite{mushe03c}.}.

As an example of the application of PIMC to the Symanzik improved action, we
consider the case of compact $U(1)$ gauge theory in three dimensions.
The relevance of the model to QCD at finite temperature 
\cite{ham94} has made it a standard proving ground for Hamiltonian
lattice numerical methods. The model has two essential 
features in common with QCD, confinement \cite{pol77,pol78,gof82,ban77}
and chiral symmetry breaking \cite{fiebig90}. Other common features are the 
existence of a mass gap and a confinement-deconfinement phase transition at 
some non-zero temperature. These similarities suggest that a comparison of 
the respective mass spectra should be informative.
In the continuum limit of theory, the mass gap $M$ is found to behave as 
\cite{pol78}
\begin{equation}
M^{2}a^{2} \approx \mbox{const.}\beta \mbox{exp}[-2\pi^{2}\beta v(0)]= M^{2}_{D}
\label{eqn10}
\end{equation}
where $v(0)\approx 0.2527$ is the lattice Coulomb Green's function at zero
separation. In 
Hamiltonian theory in which the space dimensions are discretized, 
$v_{H}(0)\approx 0.3214$ is the analogous Green's function \cite{gof82}. 
It has  been shown analytically 
that a linear confinement persists for all non-vanishing couplings, 
no matter how weak \cite{pol78,gof82}. The string tension as a function of
coupling also scales exponentially and obeys a lower bound 
\begin{equation}
K \geq \mbox{const.}M_{D}\beta^{-1}.
\label{eqn11}
\end{equation}
An interesting feature to explore in this context is whether the coupling to 
matter fields will change the permanent confinement status in (2+1) 
dimensions \cite{smorgrav03}.

\section{Method}
\subsection{Simulation details}
To extract  estimates of the string tension and the glueball masses
using the Symanzik improved action in Eq. (\ref{eqn1}), a set of simulations 
are  performed on lattices of size $N^{2}_{s}\times N_{t}$, ($N_{s}=16$, 
and $N_{t}=16-64$) where 
$N_{s}$ and $N_{t}$ are the number of lattice sites in the spatial and 
temporal directions respectively.
The lattice size in the time direction is adjusted according to the anisotropy 
used in order to keep the physical length in the spatial and temporal 
directions equal.
To analyse the behaviour in the strong and weak coupling 
regions, gauge configurations are generated using the Metropolis algorithm, 
for a range of couplings $\beta = 1-2.5$. The details of the algorithm are
discussed elsewhere \cite{mushe02}.   
Starting from an arbitrary initial gauge configuration, 50,000 sweeps 
are performed for the equilibration of the configurations and the 
self-consistent determination of the mean-field parameters. A Fourier 
acceleration procedure \cite{batrouni85,davis90} is used to overcome the 
stiffness against variations in the temporal plaquettes for high 
anisotropies. About 50$\%$ of the ordinary Metropolis updates are replaced 
with Fourier updates for $\xi\leq 0.4$ and about 100,000 further 
sweeps are 
performed to allow the system to equilibrate. 

After thermalization, 
configurations are stored every 300 sweeps; 1200 stored gauge configurations
are used in the measurement of the static quark potential and string 
tension and 1500 
configurations for the glueball masses at each $\beta$. Measurements made on 
the stored configurations  are binned into five blocks with each block 
containing an average of 250 measurements. The  mean and standard deviation 
of the final observables are estimated simply by averaging over the block 
averages.
The simulation parameters used for each configuration set are shown in 
Table \ref{tabpara}.

\begin{table}[!h]
\caption{
\label{tabpara}
Simulation parameters at various $\beta$ and $\xi$ vales.} 
\begin{ruledtabular}
\begin{tabular}{ccccccc}
Volume & $\beta$ & $\xi$ & $u_{t}$ &  $u_{s}$ & $(u_{s})^{4}$ & 
$<P> \mbox{at}\hspace{0.04cm} u_{s}$ \\  \hline
$16^{2}\times 32$ & 1.0   & 0.5  & 0.9991(3) & 0.9218(2) & 0.7220 &0.7221(4)\\
$16^{2}\times 32$ & 1.45  & 0.5  & 0.9991(2) & 0.9468(2) & 0.8038 &0.8039(3)\\
$16^{2}\times 32$ & 1.75  & 0.5  & 0.9997(3) & 0.9551(3) & 0.8324 &0.8325(6)\\
$16^{2}\times 32$ & 2.0   & 0.5  & 1.      & 0.9589(2)  & 0.8458 & 0.8459(4)\\
$16^{2}\times 32$ & 2.5   & 0.5  & 1.      & 0.9614(4)  & 0.8543 & 0.8544(3)\\
$16^{2}\times 40$ & 1.0   & 0.4  & 1.      & 0.9182(3)  & 0.7110 & 0.7111(4)\\
$16^{2}\times 40$ & 1.45  & 0.4  & 1.      & 0.9432(3)  & 0.7917 & 0.7918(7)\\
$16^{2}\times 40$ & 1.75  & 0.4  & 1.      & 0.9497(3)  & 0.8138 & 0.8139(5)\\
$16^{2}\times 40$ & 2.0   & 0.4  & 1.      & 0.9524(5)  & 0.8228 & 0.8229(5)\\
$16^{2}\times 40$ & 2.5   & 0.4  & 1.      & 0.9561(2)  & 0.8358 & 0.8359(2)\\
$16^{2}\times 48$ & 1.0   & 0.333  & 1.    & 0.9152(3)  & 0.7016 & 0.7017(5)\\
$16^{2}\times 48$ & 1.45  & 0.333  & 1.    & 0.9356(3)  & 0.7663 & 0.7664(4)\\
$16^{2}\times 48$ & 1.75  & 0.333  & 1.    & 0.9401(2)  & 0.7815 & 0.7816(6)\\
$16^{2}\times 48$ & 2.0   & 0.333  & 1.    & 0.9408(4)  & 0.7836 & 0.7837(4)\\
$16^{2}\times 48$ & 2.5   & 0.333  & 1.    & 0.9454(4)  & 0.7991 & 0.7992(3)\\
$16^{2}\times 64$ & 1.0   & 0.25  & 1.    & 0.9112(6)  & 0.6895 & 0.6892(3)\\
$16^{2}\times 64$ & 1.45  & 0.25  & 1.    & 0.9287(3)  & 0.7440 & 0.7439(4)\\
$16^{2}\times 64$ & 1.75  & 0.25  & 1.    & 0.9357(5)  & 0.7665 & 0.7663(5)\\
$16^{2}\times 64$ & 2.0   & 0.25  & 1.    & 0.9381(4)  & 0.7745 & 0.7738(6)\\
$16^{2}\times 64$ & 2.5   & 0.25  & 1.    & 0.9401(4)  & 0.7812 & 0.7811(3)\\
\end{tabular}
\end{ruledtabular}
\end{table}

\subsection{The inter-quark potential and string tension}
\label{hu2:sten}
The static quark potential $V({\bf r})$ is extracted from the expectation 
values of the time-like Wilson loops $W({\bf r},t)$, which are expected to 
behave as
\begin{equation}
W({\bf r},t)\approx \sum_{i}Z_{i}({\bf r})\mbox{exp}[-tV_{i}({\bf r})],
\label{eqn12}
\end{equation}
where the summation runs over the excited state contribution to the expectation
value, and $i=1$ corresponds to the lowest energy contribution.
To obtain the optimal overlap of Wilson loop (and glueball) operators with the
lowest state, it is necessary to suppress the contamination
from excited states. This is done by using a simple APE smearing method 
\cite{morningstar97,albanese87,takahashi01} which is implemented by the
iterative replacement of the original spatial link variables by
a smeared link. Following the single-link smearing procedure, every 
space-like link variable $U_{i}(x)$ on the lattice is replaced by
\begin{equation}
U_{i}(x)  \rightarrow  P_{U(1)}\left[\alpha U_{i}(x)
+\sum_{j\neq i}\sum_{\pm}U_{j}(x)U_{i}(x\pm\hat{j})U^{\dagger}(x\pm\hat{i})
\right]
\label{eqn13}
\end{equation}  
where $U_{-i}\equiv U^{\dagger}(x-\hat{i})$ and $i$ and $j$ are purely 
spatial indices. $P_{U(1)}$ denotes the 
projection onto $U(1)$ and $\alpha$ is the 
smearing parameter. Operators constructed out 
of smeared links dramatically reduced the mixing with high frequency 
modes, thus removing the excited-state contamination in the correlation 
functions. The smearing fraction  is fixed to $\alpha=0.7$ and ten iterations 
of the smearing process are used.   
To  reduce the variance, we use the technique of 
\emph{thermal averaging} \cite{nogueira00,mushe02}, which 
amounts to replacing the time-like link variables $U_{t}$ 
by their local averages. This technique was applied 
to all temporal links except those adjacent to the spatial legs of loops, 
which are not independent. The technique has a dramatic effect in reducing 
the statistical noise.

\begin{figure}[!h]
\scalebox{0.45}{\includegraphics{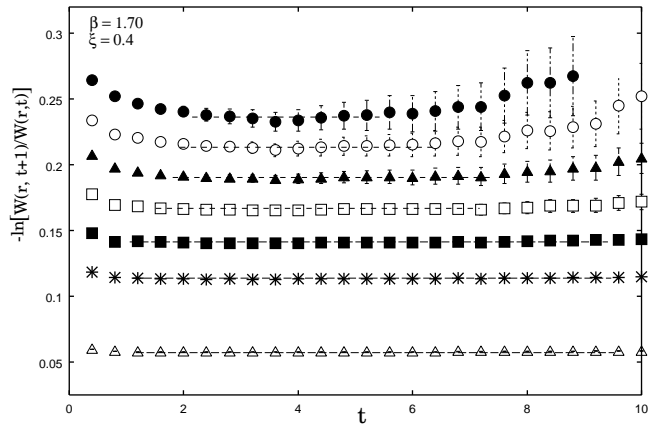}}
\caption{
\label{fig1}
Effective potential as a function of Euclidean time $t$. From bottom up
the horizantal lines correspond to $r =1$ through 7 and indicate the plateau 
values in the range $2\leq t \leq 6$.}
\end{figure}

The values of the effective potential are measured 
from the logarithmic ratio of successive Wilson loops
\begin{equation}
V_{t}({\bf r})= -\mbox{ln}\left[\frac{W({\bf r},t+1)}{W({\bf r},t)}\right]
\label{eqn14}
\end{equation}
which is expected to be independent of $t$ for $t>0$. A plot of the effective
potential is shown in Fig. \ref{fig1} for $\beta =1.70$ and $\xi = 0.4$ 
for various separations $\bf r$. The dashed lines indicate the plateau 
values at various separations. As a result of heavy smearing, a good plateau 
behaviour is seen at small $t$ values for $r =1$ through 7. For
$ r\geq 6$, we see that the signal is dominated by noise for $t>6$. 
We fix the fitting range to be, in most cases, $t=2$ to 6.

The string tension is then extracted  from the Wilson loops by establishing 
the  linear behaviour for the static quark potential at large separation 
${\bf r}$.
We have chosen to fit our results for $V({\bf r})$ to the form \cite{gof82}
\begin{equation}
V({\bf r})=a+K{\bf r}+b\mbox{ln}(\bf r),
\label{eqn15}
\end{equation}
where the linear term dominates the behaviour at large 
separations and a logarithmic Coulomb term dominates at small separations. 

\subsection{Renormalization of Anisotropy}
Since the anisotropy ratio, $\eta$, is important in QCD simulations on the 
anisotropic lattices, we study its behaviour by numerical simulations.
Measurements of renormalization of anisotropy \cite{norman98} have been made 
by comparing the static quark potential extracted from correlations along the 
different spatial and temporal directions. On an anisotropic 
lattice there are two different potentials $V_{t}(r)$ and $V_{s}(r)$
extracted from two different types of loops: time-like 
$W_{xt}$ and space-like $W_{xy}$ Wilson loops. 
The two potentials differ by a factor of $\xi_{phys}$ and by an additive 
constant, since the self-energy corrections to the static quark potential 
are different if the quark and anti-quark propagate along the temporal or 
a spatial direction. The natural way to proceed is to build ratios of the 
Wilson loops.  
\begin{eqnarray}
R_{t}(x,t)& =& \frac{W_{xt}(x,t+1)}{W_{xt}(x,t)}\nonumber\\
R_{s}(x,y)& = &\frac{W_{xy}(x,y+1)}{W_{xy}(x,y)}
\label{eqn16}
\end{eqnarray}
Asymptotically for large $\tau$ and $y$, the ratios $R_{t}$ and $R_{s}$
approach
\begin{eqnarray}
R_{t}(x,t)\approx Z_{x\tau}\mbox{e}^{-\tau V_{t}}+(\mbox{excited state 
contr.})\nonumber\\
R_{s}(x,y)\approx Z_{xy}\mbox{e}^{-yV_{s}}+(\mbox{excited state contr.})
\label{eqn18}
\end{eqnarray}
The physical anisotropy is then determined from the ratio of the 
potentials $V_{t}(r)$ and $V_{s}(r)$ estimated from $R_{t}$ and $R_{s}$ 
respectively. 
The lattice potentials defined by Eq. (\ref{eqn18}) contain contributions 
from the self-energy terms. The potential is simply parameterized as
\begin{equation}
V_{s}(\xi ,r) = V^{0}_{s}(\xi )+V^{f}_{s}(\xi ,r),
\label{eqn19}
\end{equation}
where $V^{f}_{s}$ is the lattice potential free of self-energy 
contributions. The time-like potential $V_{t}$ is treated similarly. To 
eliminate the effect of the self-energy term $V^{0}$ in the potentials, we
define a subtracted potential
\begin{eqnarray}
V^{sub}_{s}(\xi ,r,r_{0}) & =& V^{f}_{s}(\xi ,r)-V^{f}_{s}(\xi ,r_{0})
\nonumber\\
V^{sub}_{f}(\xi ,t,t_{0}) & =& V^{f}_{t}(\xi ,t)-V^{f}_{t}(\xi ,t_{0})
\label{eqn20}
\end{eqnarray}
where the subtraction points $r_{0}$ and $t_{0}$ are chosen to satisfy
$t_{0} = \xi r_{0}$ and the matching of the potential
$V^{f}_{t}(t_{0}=\xi r_{0})=V^{f}_{s}(r_{0})$ should be satisfied there. 
The subtraction radius $r_{0}$ should be chosen to be as small as possible 
so that fluctuations of the potential does not increase in which case 
simulations with high statistics on a larger lattice are required. 
The renormalized anisotropy is determined from the ratio
\begin{equation}
\xi_{phys} = \frac{V^{sub}_{t}(\xi ,r,r_{0})}{V^{sub}_{s}(\xi ,r,r_{0})}
\label{eqn21}
\end{equation}
and the  measured anisotropy parameter, $\eta$, is then given by
\begin{equation}
\eta = \frac{\xi_{phys}}{\xi}
\label{eqn22}
\end{equation}
An alternative approach to make the comparison is to fit the measured 
potentials with the forms \cite{alford2001}
\begin{eqnarray}
a_{s}V_{s}(x) &=& a+\sigma a_{s}^{2}x+c\mbox{ln}x \nonumber\\
a_{s}V_{t}(t) &=& a+\sigma 
a_{s}a_{t}t+c\mbox{ln}\left[\frac{a_{s}}{a_{t}t}
\right]
\label{eqn23}
\end{eqnarray}
The renormalized anisotropy $\xi_{phys}$ is then determined from the
ratio of the coefficients of the linear terms in the two cases.
It is advantageous to use the potential at smaller
$r$, where the statistical errors are smaller, and determine $\xi_{phys}$ 
from the ratio of the coefficients of the Coulomb terms; however, such an 
estimate depends on short distance effects and is more sensitive to possible 
discretization errors of $O(a^{4}/r^{4})$ \cite{norman98}.

\subsection{Glueball masses}
\label{hu2:asym}
The numerical analysis of the mass of a glueball having a given $J^{PC}$
proceeds through a study of the time-like correlations between space-like
Wilson loop  operators $\Phi_{i}(t)$
\begin{equation}
C(t) = \langle\bar{\Phi}^{\dagger}_{i}(t)\bar{\Phi}_{i}(t)\rangle ,
\label{eqn24}
\end{equation}
where
\begin{equation}
\bar{\Phi}_{i}(t)= \Phi_{i}(t)-\langle 0\mid\Phi_{i}(t)\mid 0\rangle
\label{eqn25}
\end{equation}
is a gauge invariant vacuum subtracted operator capable of creating a 
glueball out of the vacuum. It is necessary to to subtract the vacuum 
contribution for the scalar glueball with $J^{PC}=0^{++}$, 
because in the large Euclidean time limit, $C(t)$ becomes  
dominated  by the lowest energy state carrying the quantum numbers of $\Phi$
and these quantum numbers may coincide with that of the vacuum. The vacuum 
contribution is averaged over the whole ensemble before subtracting from 
the correlator. The glueball mass of interest is then extracted by 
studying the exponential decay of the correlation function for large 
Euclidean times, which is expected to behave as
\begin{eqnarray}
C(t)& =& c_{i}\left[\mbox{exp}(-m_{i}t)+\mbox{exp}(-m_{i}(T-t))\right]
\nonumber\\
& & +(\mbox{excited state contributions})
\label{eqn26}
\end{eqnarray}
where $m_{i}$ is the mass of the lowest-lying glueball which can 
be created by $\bar{\Phi}_{i}(t)$, and $T=N_{t}a_{t}$ is the extent of the 
periodic lattice in the time direction. Here, only the lowest
 ``symmetric" ($PC=++$) and ``antisymmetric" ($PC= --$) glueball states are
studied. The measured values of $C(t)$ are expected to fall on a simple 
exponential curve assuming that the lattice is fine enough for the glueball 
mass to exhibit  scaling behaviour according to the theoretical predictions. 

On a finite lattice with lattice spacing $a$, the operator $\Phi_{i}(t)$ 
has a small overlap with the glueball ground state and the  mass 
extracted from $C(t)$ may be too large owing to the excited-state 
contamination. 
The overlap gets worse as the lattice spacing is reduced and nears
 the continuum limit, $a\rightarrow 0$. This is obvious since 
the physical extension of the glueball remains fixed, while the operator 
$\Phi$, constructed from small loops, probes an ever smaller region of the 
glueball wave function as the lattice spacing is decreased. 
Hence it becomes important to use an improved  glueball operator so as to 
have approximately the same size as the physical size of the glueball. For 
such an operator, the overlap with the glueball of interest is strong 
at small lattice spacing and signal-to-noise ratio is also optimal 
\cite{morningstar96}.

Following the variational technique of Morningstar 
and Peardon \cite{morningstar97} and the smearing procedure of Teper 
\cite{tep99}, an optimized operator was found as a linear combination of the
basic operators $\phi$
\begin{equation}
\Phi(t)=\sum_{\alpha}v_{i\alpha}\phi_{i\alpha}(t)
\label{eqn27}
\end{equation}
where the index $\alpha$ runs over the rectangular Wilson loops with 
dimensions $l_{x}=[n-1, n+1]$, $l_{y}=[n-1, n+1]$ and smearing 
$n_{s}=[m-1, m+1]$.  
The correlation function $C(t)$ is then computed from the optimized 
glueball operator $\bar{\Phi}_{i}(t)$
\begin{equation}
C_{i}(t) =\sum_{t_{0}}\langle 0\mid 
\bar{\Phi}_{i}(t+t_{0})\bar{\Phi}_{i}(t_{0})\mid 0\rangle
\label{eqn28}
\end{equation}

\begin{figure}[!h]
\scalebox{0.45}{\includegraphics{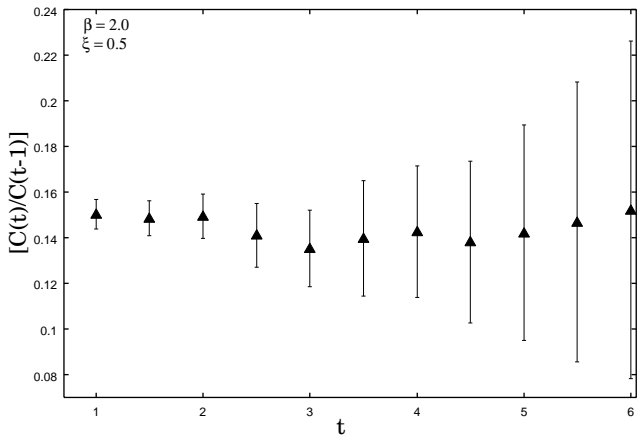}}
\caption{
\label{fig2}
Effective mass plot for the scalar glueball at $\beta = 2.0$ and $\xi 
=0.50$.}
\end{figure}

Fig. \ref{fig2} shows the effective mass plot for the scalar glueball for 
the measurements at $\beta = 2.0$ and $\xi = 0.50$. The signal is seen out 
to the time slice 6 and reaches a plateau  region for $1\leq t \leq 5$. 
The data are noisy for $t>5$.

We choose to fit the correlation function $C(t)$ with the 
simple form
\begin{equation}
C_{i}(t ) = c_{1}\cosh m_{i}[T/2 -t]
\label{eqn29}
\end{equation}
to determine the glueball mass estimates.

\section{Results and discussion}

\subsection{Static quark potential and rotational symmetry}
\begin{figure}[!h]
\scalebox{0.45}{\includegraphics{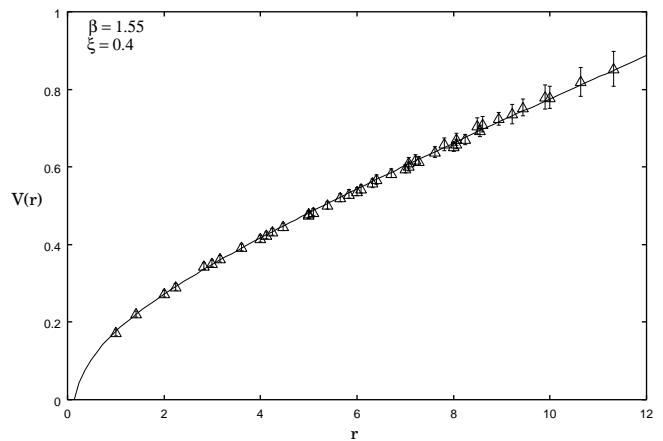}}
\caption{
\label{fig3}
Static potential, $V(r)$, as a function of separation $r$ at $\beta =1.55$ 
and $\xi = 0.4$. The solid line is a fit to the form $V(r)=a+br+c\mbox{ln}(r)$ 
in the range $3\leq r\leq 8$.}
\end{figure}

A plot of the static quark potential $V(r)$ as a function of $r$ at 
$\beta =1.55$ and $\xi = 0.4$ is shown in Fig. \ref{fig3}. The data in 
this plot were obtained by looking for a plateau in the effective potential. 

Because we are concerned to make long distance behaviour consistent in both 
fine and coarse directions, it is advantageous to use Wilson loops of the 
largest possible spatial extent. However, in practice, the statistical errors 
in large Wilson loops grow exponentially with separation. We use Wilson  
loops of size $8\times 8$ and fit the data by (\ref{eqn15}). We fit in 
the range $3\leq r \leq 8$ so that we are not sensitive to the Coulomb 
term and the discretization errors associated with it.
We see that the data are fitted very well giving the string 
tension, $K(=\sigma a^{2})=0.044(3)$.

\begin{figure}[!h]
\scalebox{0.45}{\includegraphics{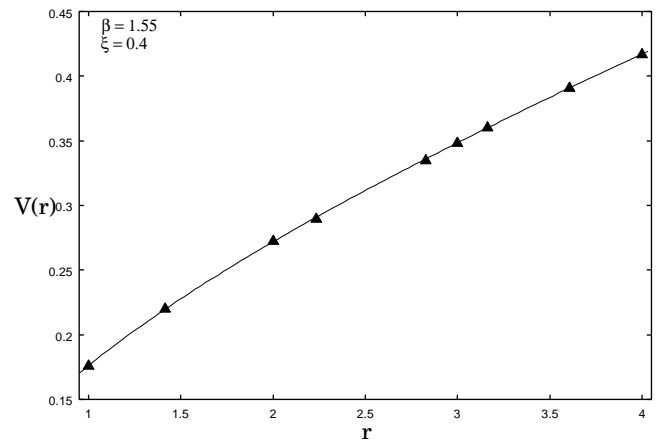}}
\caption{
\label{fig4}
A close up of the static quark potential, $V(r)$, at small $r$, using the 
Symanzik improved action. The solid line is a fit of Eq. (\ref{eqn15}) 
to on-axis points $r= 4$ to 8. This plot involves  measurements at $\beta 
= 1.55$ for $\xi =0.4$ with 10 smearing sweeps at smearing parameter 
$\alpha =0.7$} 
\end{figure}

One of the main features of the improved discretization is the improved 
rotational invariance \cite{alford95}. Discretization errors in the gluon
action affect the extent to which continuum symmetries, such as rotational 
symmetry, are restored. To explore the extent to which 
rotational invariance  is improved, we measure the potential at off-axis as 
well as on-axis separations. Thus for improved rotational invariance, the 
static potential, for example, at ${\bf r} =(4,3)$ should agree exactly 
with that at ${\bf r}= (5,0)$. 

\begin{figure}[!h]
\scalebox{0.45}{\includegraphics{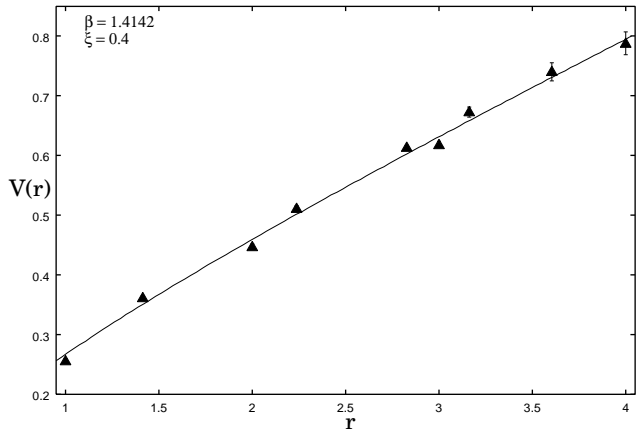}}
\caption{
\label{fig5}
A close up of the static quark potential, $V(r)$, at small $r$. The solid 
line 
is a fit of Eq. (\ref{eqn15}) to on-axis points $r= 4$ to 8.
This plot involves  measurements from the standard Wilson action 
\protect\cite{mushe02} at 
$\beta = 1.4142$ for $\xi =0.4$ with 10 smearing sweeps at smearing parameter 
$\alpha =0.7$.} 
\end{figure}

For a fixed number of configurations and constant physical volume, we 
show the results from the Symanzik improved action and standard Wilson 
action in Figs. \ref{fig4} and \ref{fig5} respectively. We see that the off 
axis points for the improved lattice are excellently fitted by the 
rotationally invariant fitting curve (\ref{eqn15}) through $r =3$ to 8. 
The data from the Wilson action lie rather less close to the line of 
best fit. However, at large separations, the standard Wilson action does 
just as well as  the improved action in extracting the static quark 
potential.

As a quantitive measurement of the improvement, the potential measured in 
the simulation from non-planar Wilson loops is compared with an 
interpolation to the on-axis data \cite{norman98}:
\begin{equation}
\Delta V(r)\equiv \frac{V_{\mbox{sim}}(r)-V_{\mbox{fit}}(r)}{\sigma r}.
\label{eqn30}
\end{equation}
Results for ${\bf r} = (1,1)$ are given in Table \ref{tab2}. With the 
mean-field inspired Symanzik improvement, the difference is only a few percent 
compared to a difference of about 10-20\% for the Wilson action \cite{mushe02}.

\begin{table}[!h]
\caption{
\label{tab2}
Comparison between the measured anisotropy $\xi_{phys}$ and the input 
anisotropy $\xi$ for the Symanzik improved and the standard Wilson 
actions \protect\cite{mushe02}. The measured difference in the off-axis 
potential at ${\bf r}=(1,1)$ are also shown.}
\begin{ruledtabular}
\begin{tabular}{cccccc}
Action &$\beta$ & $\xi$  & \multicolumn{2}{c}{$\xi_{phys}$} & 
$\Delta V(\sqrt{2})$ \\ 
& & & $r_{0}=2$ & $r_{0}=\sqrt{2}$ & \\ \hline 
Improved & 1.35  & 0.50  & 0.496(5)& 0.493(2)  & 0.03(1)\\
action   &       & 0.40  & 0.392(4)& 0.390(6)  & 0.03(2)\\
         &       & 0.333 & 0.328(3)& 0.320(5)  & 0.04(4)\\  
         & 1.45  & 0.444 & 0.442(6)& 0.441(4)  & 0.04(2)\\
         &       & 0.333 & 0.326(3)& 0.321(6)  & 0.04(3)\\
         &       & 0.25  & 0.246(4)& 0.240(7)  & 0.05(4)\\ 
         & 1.55  & 0.40  & 0.402(2)& 0.398(6)  & 0.04(2)\\
         &       & 0.25  & 0.241(5)& 0.239(7)  & 0.06(4)\\
         & 1.65  & 0.333 & 0.334(5)& 0.332(5)  & 0.02(1)\\
         & 1.75  & 0.25  & 0.252(4)& 0.249(6)  & 0.03(1)\\
         & 2.0   & 0.333 & 0.335(6)& 0.334(2)  & 0.03(1)\\
         &       & 0.25  & 0.242(4)& 0.246(1)  & 0.06(2)\\     
Wilson   & 1.35  & 0.444 & 0.418(5)& 0.419(7)  & 0.12(1)\\
action   & 1.55  & 0.40  & 0.379(3)& 0.374(6)  & 0.10(1)\\
         & 1.70  & 0.333 & 0.286(7)& 0.289(5)  & 0.08(1)\\
         & 2.0   & 0.333 & 0.281(6)& 0.292(2)  & 0.13(2)\\ 
\end{tabular}
\end{ruledtabular}
\end{table} 
 
\subsection{Numerical determination of renormalized anisotropy}

\begin{figure}[!h]
\scalebox{0.45}{\includegraphics{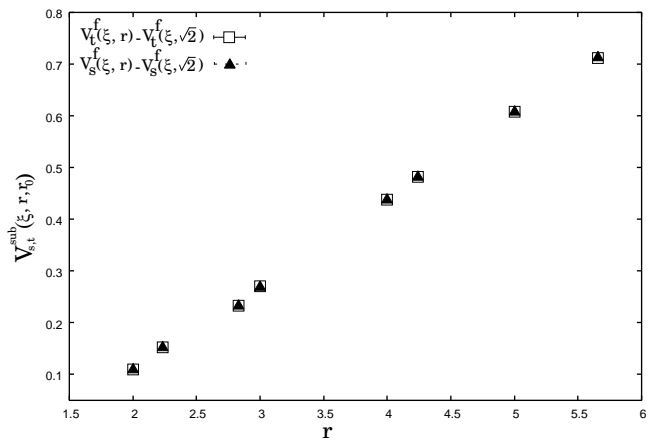}}
\caption{
\label{fig6}
Subtracted spatial and temporal potentials against the separation $r$. 
This plot involves measurements at $\beta = 1.75$ and $\xi = 0.333$ for the 
subtraction point $r_{0}=\sqrt{2}$. The temporal potential $V^{sub}_{t}$ has 
been rescaled by the input anisotropy.}
\end{figure}

We choose the points $r_{0}=2$ and $\sqrt{2}$ to compute the subtraction 
potentials of Eq. (\ref{eqn20}) and use them to obtain the ratio in 
Eq. (\ref{eqn21}). The subtracted spatial and temporal 
potentials at the subtraction point $r_{0}=\sqrt{2}$ are shown in 
Fig. \ref{fig6}. The anisotropies measured at these 
subtraction points for the improved and unimproved actions at different 
$\beta$ values are compared in Table \ref{tab2}. The two determinations of the 
anisotropies are in excellent agreement. These results show that the input 
anisotropy is normalized by less than a few percent for the improved action.
This is in contrast with the standard  Wilson action \cite{mushe02}, where the 
measured anisotropy is found to be about 20-30\% lower than the input 
anisotropy $\xi$.

\begin{figure}[!h]
\scalebox{0.45}{\includegraphics{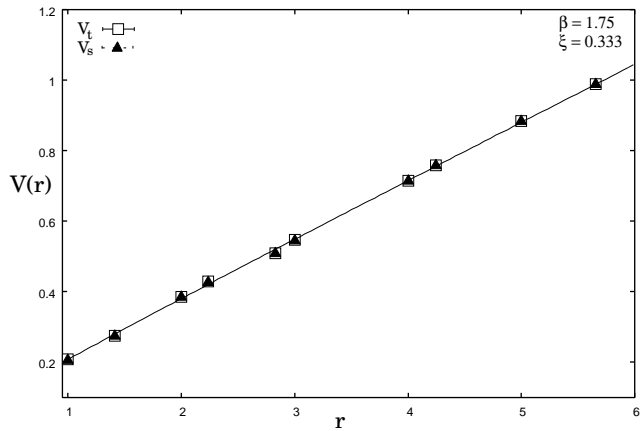}}
\caption{
\label{fig6a}
Unsubtracted spatial and temporal potentials against the separation $r$ at  
$\beta = 1.75$ and $\xi = 0.333$. The temporal potential $V_{t}(r)$ 
has been rescaled by the input anisotropy.}
\end{figure}

Fig. \ref{fig6a} shows the potentials computed from spatial and temporal 
Wilson loops without subtracting the self energy terms. We find that the 
difference between the estimates computed from subtracted and unsubtracted
potentials  is less than 1\% for the Symanzik improved action.
We conclude that a few percent renormalization in the anisotropy 
is sufficiently small that it is unlikely to represent the dominant 
effect in the final estimates and it is safe to use the bare anisotropy 
for the tadpole improved Symanzik action.

\subsection{String tension}

\begin{figure}[!h]
\scalebox{0.45}{\includegraphics{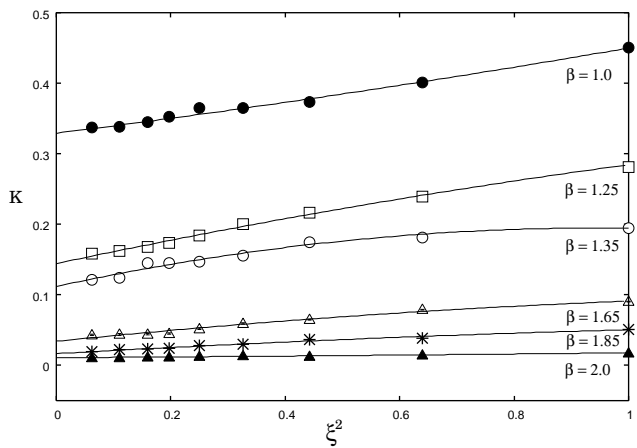}}
\caption{
\label{fig7}
Extrapolation of the string tension to the Hamiltonian limit 
$\xi\rightarrow 0$, for various $\beta$. Solid lines show the 
quadratic fits in $\xi^{2}$ to the data.}
\end{figure}

To obtain estimates of the string tension in the Hamiltonian limit, 
an extrapolation is performed by a simple quadratic fit in  powers of 
$\xi^{2}$ for each  $\beta$ value. The simulations run over a range of 
anisotropies, $\xi= 1-0.25$, thus enabling reliable extrapolation to the 
Hamiltonian limit. The errors for the extrapolation may be obtained by the 
``linear, quadratic, cubic" extrapolant method \cite{per00}. 
Fig. \ref{fig7} shows our estimates of the string tension as a function 
of the anisotropy $\xi^{2}$ for various fixed $\beta$ values. Except at 
$\beta = 1.35$, a fairly smooth variation of string tension with 
$\xi^{2}$  for various couplings  is seen. The curvature in the 
extrapolation at $\beta = 1.35$ suggests that our estimate may be 
somewhat  too high there.      

\begin{figure}[!h]
\scalebox{0.45}{\includegraphics{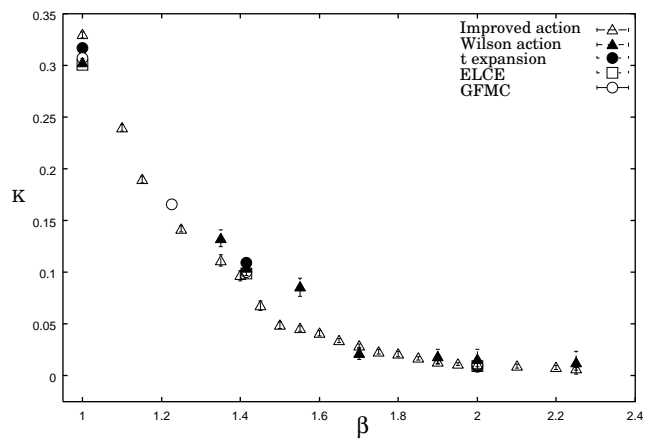}}
\caption{
\label{fig8}
The string tension as a function of inverse coupling. 
Our present estimates are shown as open triangles.
Earlier results from standard Wilson action \cite{mushe02}, 
$t$-expansion \protect\cite{mor92}, Greens Function Monte Carlo simulations 
\protect\cite{ham00} and an exact linked cluster expansion 
\protect\cite{irv84} are shown as solid triangles, solid circles, 
open circles and open squares respectively.}
\end{figure}

Our extrapolated results for the string tension, $K=a^{2}\sigma$, together 
with the earlier
Hamiltonian estimates obtained from the $t$-expansion \cite{mor92}, 
Green's Function Monte Carlo simulations \cite{ham00} 
and the Exact Linked Cluster Expansion (ELCE) \cite{irv84} are plotted as 
a function of inverse coupling in Fig. \ref{fig8}. 
We see that the string tension displays an exponential behaviour at  
weak coupling in accordance with the theoretical prediction.
It has rigorously been shown that the string tension in $U(1)_{(2+1)}$ 
undergoes a  `\emph{roughening transition}'  \cite{irv83,kogut81} at some 
intermediate coupling  estimated to be near $\beta \approx 0.8$. Beyond 
the transition point, the different estimates of  the string tension are 
expected to agree. The on-axis strong coupling series approximants fail to 
converge beyond $\beta =0.8$, which prevents the analytic continuation of 
the series expansion beyond the roughening  transition. The $t$-expansion 
results, however, do not suffer from this difficulty. A comparison with the 
GFMC \cite{ham00} and an exact linked-cluster expansion \cite{irv84} and 
results shows that our estimates are in good agreement with earlier 
estimates. The $t$-expansion estimates \cite{mor92} are a little high, 
but still reasonably accurate.
Of course, we do not expect that the results for the improved action should
match exactly at finite coupling with other estimates which were computed for 
the unimproved action. We believe our PIMC estimates are more 
reliable and accurate and are also clearly consistent with  the behaviour 
predicted by Polyakov \cite{pol78} and G{\" o}pfert and Mack \cite{gof82}.

\begin{figure}[!h]
\scalebox{0.45}{\includegraphics{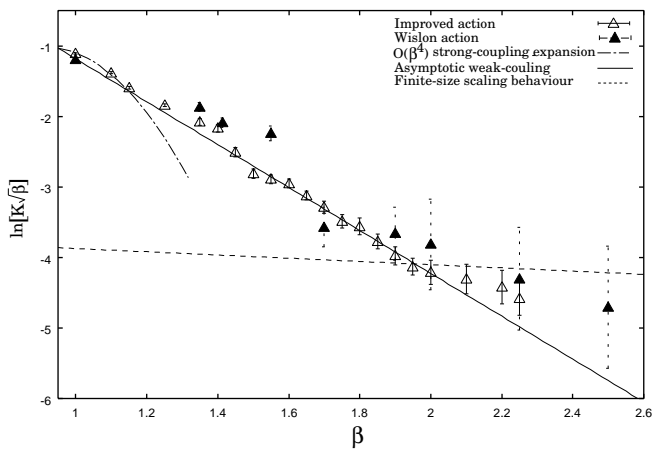}}
\caption{
\label{fig9}
The logarithm of the string tension as a function of inverse 
coupling. 
Our present estimates are shown as open triangles.
The solid line is the result of fitting to  form 
(\protect\ref{eqn6}) for $1.45\leq \beta \leq 1.95$. The dash-dot line is the
$\beta^{4}$ order strong coupling expansion \protect\cite{hamer96a}. Our 
previous estimates \protect\cite{mushe02} are shown as solid triangles. 
The dashed line represents the finite size behaviour \protect\cite{ham93}.}
\end{figure}

Fig. \ref{fig9} shows the scaling behaviour of the string tension 
together with  results obtained using the standard Wilson action 
\cite{mushe02},  as a function of $\beta$. 
The dashed-dot line is the strong-coupling expansion to order $\beta^{4}$ 
\cite{hamer96a} and the solid line represents a fit to the weak-coupling 
asymptotic form (\ref{eqn6}). 
It can be seen that our present estimates appear to match nicely onto the 
strong and weak
coupling expansions in their respective limits.
The $4th$ order strong coupling series expansion, 
obtained from integrated differential approximants, diverges beyond 
$\beta =1.30$. In the weak-coupling region, the
 string tension is consistent with the 
predicted scaling behaviour \cite{gof82}.
An unconstrained fit of the form (\ref{eqn6}) represents 
the data rather well in the interval $1.45\leq \beta \leq 1.95$.
The fit to the data gives a scaling slope of $3.04\pm 0.13$ and an intercept
of $1.42\pm 0.21$. The intercept of the scaling curve is roughly two times
larger in magnitude than the theory predicts, compared to our previous 
results with the standard Wilson action \cite{mushe02} 
which were higher than theory by a factor of 5 - 6. Also in contrast with
the Wilson action, a significant reduction in the errors is clearly apparent 
with  the tadpole-improved Symanzik action.

In summary, it appears that the overall exponential scaling behaviour is the 
same for both actions, but the constant coefficient is lower for the Symanzik
action by a factor of 2 to 3, and closer to the theoretical weak-coupling 
estimate. It seems
highly plausible that a different action should give a different 
renormalization for the constant coefficient, although no analytic calculation of this effect has been done.

\subsection{Antisymmetric mass gap}

The weak-coupling behaviour of the mass gap is not exactly known. The 
rigorous analysis of G{\" o}pfert and Mack \cite{gof82} showed that in the
continuum limit $U(1)_{2+1}$ reduces to a massive scalar free field 
theory, with a mass gap $M$ which is expected to decrease exponentially as 
the lattice spacing goes to zero. They showed that the lattice photon mass 
in 
the Villain action on a 3-dimensional Euclidean lattice is given by
Eq. (\ref{eqn6}). It is often claimed in the literature that the Villain 
action is a high-$\beta$ approximation of the Wilson action so that Eq. 
(\ref{eqn6}) should also hold  in the weak-coupling limit of the 
Wilson model.

\begin{figure}[!h]
\scalebox{0.45}{\includegraphics{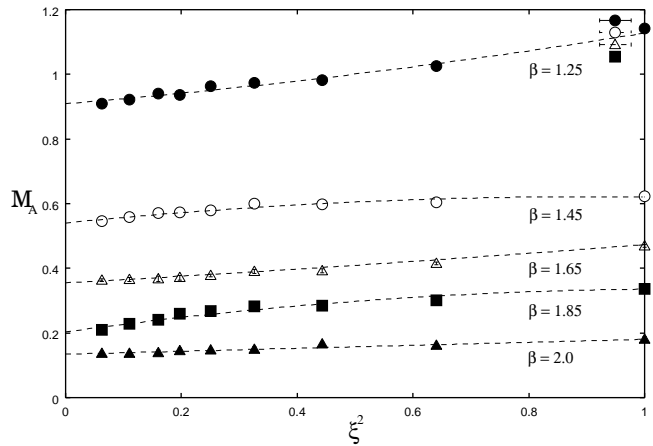}}
\caption{
\label{fig10}
Extrapolation of the antisymmetric glueball mass to the Hamiltonian 
limit $\xi \rightarrow 0$. Dashed lines show the quadratic fits in
$\xi^{2}$ to the data.}
\end{figure}

The extrapolation of the glueball masses to the Hamiltonian limit is shown 
in Fig. \ref{fig10}. The extrapolation is performed by using a simple 
quadratic fit in  powers of $\xi^{2}$. Again we see a smooth dependence 
on $\xi^{2}$, for all $\beta$ values analysed here.

\begin{figure}[!h]
\scalebox{0.45}{\includegraphics{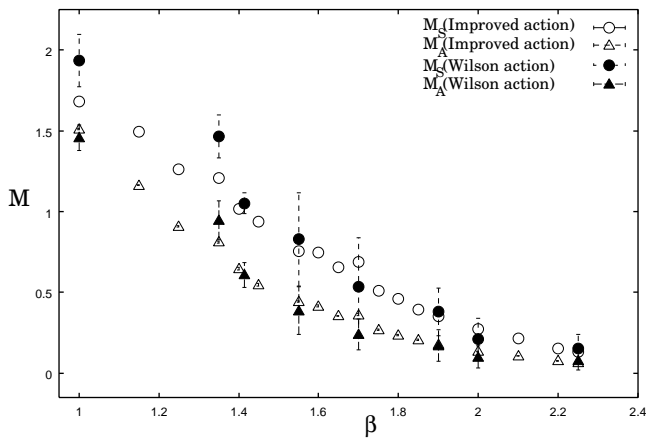}}
\caption{
\label{fig11}
Glueball masses as  functions of $\beta$. 
Our present results for the symmetric and antisymmetric glueball masses 
($M_{S}$, $M_{A}$) are shown as open circles and triangles  respectively. 
Earlier results obtained from the Wilson action \protect\cite{mushe02}
 are shown as solid circles and triangles respectively.}
\end{figure}

Our extrapolated results for the mass gap together with earlier 
Hamiltonian estimates obtained from the Wilson action \cite{mushe02} 
are plotted as a 
function of $\beta$ in Fig. \ref{fig11}. Unlike the string tension, 
the strong-coupling expansion of the  mass gaps is believed to be analytic 
near the roughening point. Comparison with earlier Hamiltonian estimates 
shows that  our present data follow quite closely the strong-coupling 
expansion estimates, obtained by the method of integrated 
differential approximants, in the strong and weak-coupling regions. 
The $t$-expansion estimates \cite{mor92}, obtained from D-Pad{\'e} 
approximants, are found to be substantially less accurate 
than those from the SC-expansion. Beyond $\beta =2$, 
the approximants for the t-expansion do not converge well 
so that no reliable estimates can be obtained beyond that coupling 
value. 

\begin{table}[!h]
\caption{
\label{tadres}
Comparison of the Hamiltonian results obtained for the 
symmetric and antisymmetric  scalar glueball masses ($M_{S}$, 
$M_{A}$) for the Symanzik improved and Wilson actions.}

\begin{ruledtabular}
\begin{tabular}{ccccc} 
$\beta$ &\multicolumn{2}{c}{Improved action}&\multicolumn{2}{c}{Wilson action}
\\
  & $M_{S}$ & $M_{A}$ & $M_{S}$ & $M_{A}$  \\ \hline
1.0  &1.68(1)  &1.510(5) & 1.9(1)  & 1.45(7)  \\
1.35 &1.207(9) &0.811(5) & 1.4(1)  & 0.9(1)    \\
1.55 &0.75(1)  &0.444(5) & 0.8(2)  & 0.4(1)     \\ 
1.70 &0.69(2)  &0.361(6) & 0.5(3)  & 0.24(9)    \\
1.90 &0.352(6) &0.177(3) & 0.3(1)  & 0.17(9)    \\
2.0  &0.272(4) &0.136(2) &0.2(1)   &0.10(6)    \\
\end{tabular}
\end{ruledtabular}
\end{table}  

\begin{figure}[!h]
\scalebox{0.45}{\includegraphics{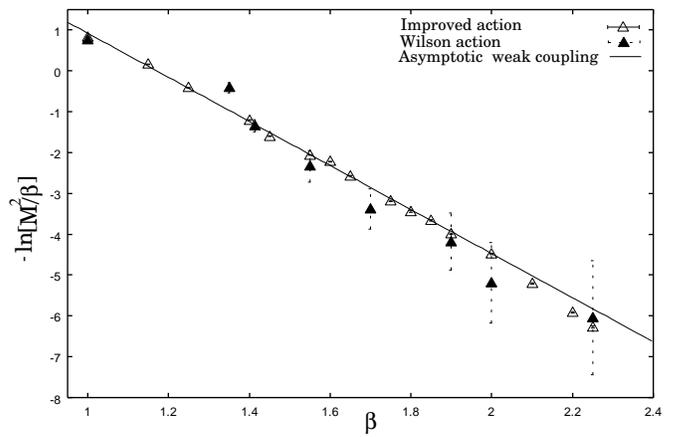}}
\caption{
\label{fig12}
The logarithm of the antisymmetric mass gap as a function of $\beta$.
Open triangles show our present estimates.
The solid line is the fit to the data for $1\leq \beta \leq 2$. 
Previous estimates from the Wilson action \protect\cite{mushe02}
 are shown as  solid triangles.}
\end{figure}

The asymptotic scaling behaviour of the antisymmetric mass gap is shown in 
Fig. \ref{fig12}. The solid line is the result of fitting for $1.4\leq 
\beta \leq 2.25$ to the form (\ref{eqn10}) to find the
scaling slope and the intercept of the scaling curve. 
Our results for these  coefficients are shown and compared  with previous 
studies  in Table \ref{tab:massgap}. These results are obtained by fitting to
 the form $M^{2} = \beta \mbox{exp}[-f_{0}\beta +f_{1}]$ in the 
weak-coupling region. It can be seen that the agreement with the earlier 
results is remarkable. We find the constant coefficient (intercept of the 
scaling curve) is approximately 1.5 times larger in magnitude than the 
theory predicts. This is an improvement over our 
previous estimate  using the Wilson 
action  \cite{mushe02}, where the constant coefficients were estimated
a factor 5-6 times larger. The scaling slope is a little 
less than the theoretical prediction but in agreement with the estimates
obtained in other numerical and  analytic calculations (Ref. Table 
\ref{tab:massgap}). Several studies have provided evidence that the
antisymmetric mass gap in the Wilson model of $U(1)_{2+1}$ Hamiltonian 
lattice gauge theory does not fall in the weak-coupling limit in the same 
manner as the periodic Gaussian model\footnote{Based on the fact that periodic 
Gaussian models are special forms of Wannier-fuction expansions, Suranyi 
\protect\cite{suranyi83} has argued 
that a natural series of models, beginning with periodic Gaussian and 
approximating the Wilson model with arbitrary precision, does not exist.}.
This may be a signal of non-universality for the Abelian $U(1)$ theory.

\begin{table}[!h]
\caption{
\label{tab:massgap}
Results for the coefficients $f_{0}$ (slope of the scaling curve)
 and $f_{1}$ (the intercept of the scaling curve) in the weak-coupling
 limit for the antisymmetric mass gap.}
\begin{ruledtabular}
\begin{tabular}{ccc}
Source & $f_{0}$ & $f_{1}$  \\ \hline
Villain (Hamiltonian) \protect\cite{gof82}  & 6.345 & 4.369\\
Morningstar \protect\cite{mor92} & 5.23 & 5.94\\
Hamer and Irving \protect\cite{hamer85a} & 5.30 & 6.15\\
Hamer, Oitmaa and Weihong \protect\cite{ham92} & 5.42 & 6.27\\
Plaquette Expansion \protect\cite{john97} & 5.01 & 5.82\\
Heys and Stump \protect\cite{hey85} & 4.97 & 6.21\\
Lana \protect\cite{lana88} & 4.10 & 4.98\\
Xiyan, Jinming and Shuohong \protect\cite{fan96} & 5.0 & 5.90\\
Dabringhaus, Ristig and Clark \protect\cite{dab91} & 4.80 & 6.26\\
Darooneh and Modarres \protect\cite{dar2000} & 4.40 & 5.78\\
Present Work & 5.39(9) & 6.3(1)\\
\end{tabular}
\end{ruledtabular}
\end{table}

\subsection{Mass gap ratio}
The quantity of interest here is the dimensionless ratio $R_{M}$ 
between the symmetric and antisymmetric mass gap in the large $\beta$ limit. 
In this limit the ratio $R_{M}=M_{S}/M_{A}$ is expected to tend 
smoothly to its continuum value. In practice, this limiting value is found
by increasing $\beta$ from strong-coupling until the mass ratio levels off in
the weak coupling region. Earlier studies of the photon mass 
\cite{mor92} showed that the scaling of the mass ratio sets in 
for $\beta >1$. If the continuum theory admits a stable bound state of two 
photons (a glueball), then the weak-coupling limit of the mass ratio will 
lie between 1 and 2. If the continuum theory is simply a free-field theory 
of massive scalar photons, as in the Villain model or if the glueball 
remains in the continuum theory only as a resonance, then the ratio 
$R_{M} =2$ should be observed as $\beta$ becomes large. Lana \cite{lana88} 
argued for a sharp ``transition point" at $\beta = 1.40$ where the symmetric 
bound state ceases to be stable, and crosses the level consisting of two free 
axial particles, so that mass ratio suddenly levels out at
value 2.   

\begin{figure}[!h] 
\scalebox{0.45}{\includegraphics{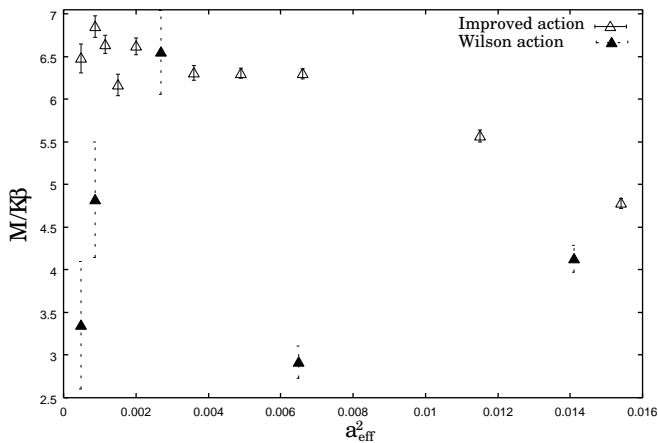}} 
\caption{
\label{fig13}
The dimensionless ratio $M_{A}/\beta K$ as a function 
of effective lattice spacing, $a_{eff}$. Open 
triangles show our present PIMC estimates. 
Our previous estimates using the Wilson action \protect\cite{mushe02}
 are shown as solid triangles.}
\end{figure}

In general, the most accurately calculated physical quantity is the string 
tension $K$. Therefore the first quantity that we shall extrapolate to the 
continuum limit will be of the form $M/K$, where $M$ is the glueball mass. 
The leading corrections to such a ratio are known to be of the order
$O(1/\zeta)^{2}$ \cite{tep99}, where $\zeta$ is some length scale.
In the weak-coupling limit,
\begin{displaymath}
M_{A}/K \rightarrow \pi^{2}/2 \hspace{0.40cm} \mbox{as} \hspace{0.30cm}
\beta \rightarrow \infty ,
\end{displaymath}
in constrast to  QCD, where $M/\sqrt{K}$ is constant in the continuum limit. 
Fig. \ref{fig13} displays the estimates for this quantity, as a function 
of $\beta$. The data certainly appear to level out below $a^{2}_{eff} \approx 
0.006$  at a value 6. This is quite close to  the predicted value 
for the Villain model.

\begin{figure}[!h]
\scalebox{0.45}{\includegraphics{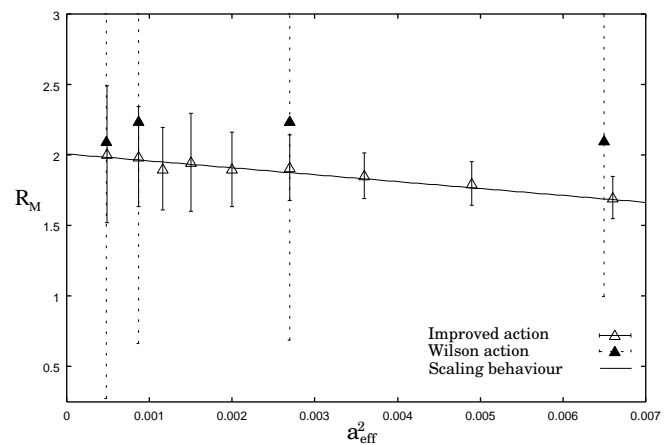}}
\caption{
\label{fig14}
Mass ratio, $R_{M}$, as a function of effective lattice 
spacing, $a_{eff}$.
Our present estimates are shown as the open triangles. 
The solid line is the 
fit to the data in the range $0.00026\leq a^{2}_{eff}\leq 0.0066$.
The solid triangles show our earlier estimates obtained from the
Wilson action \protect\cite{mushe02}.}
\end{figure}

Figure \ref{fig14} shows the behaviour of the dimensionless mass ratio 
 $R_{M}=M_{S}/M_{A}$ as 
a function of effective lattice spacing squared, $a^{2}_{eff}$, where
$a_{eff}$ is defined from Eq. (\ref{eqn10}) as \cite{mushe02}
\begin{displaymath}
a_{eff} =\sqrt{8\pi^{2}\beta}e^{-\pi^{2}\beta v(0)}.
\end{displaymath}
 The plot 
shows that the mass ratio approaches very 
closely to  the expected value of 2 in the large $\beta$ limit. 
A linear fit to the data from $0.00026\leq a^{2}_{eff} \leq 0.0066 $ gives 
$R_{M}=2.007\pm0.01$, which is 
consistent with a continuum limit value $R_{M}=2$. 
There is no sign of a stable, scalar glueball bound state,  nor of any 
sharp "\emph{break}" in the mass ratio. In comparison with previous results
using the unimproved action, there are two notable features. 
First, the results are much more accurate. Secondly, the corrections to the
continuum limit appear to be linear in $a^{2}_{eff}$, whereas those for the 
unimproved action were linear in $a_{eff}$ \cite{mushe02}. This provides 
impressive evidence that the improved action has achieved what it was 
designed to do: produce faster convergence to the continuum limit.
\begin{table}[!h]
\caption{
\label{tab:results}
Comparison of the Hamiltonian estimates for dimensionless mass ratio $R_{M}$
obtained using the improved and standard Wislon actions}
\begin{ruledtabular}
\begin{tabular}{ccc} 
$\beta$ &\multicolumn{2}{c}{$R_{M}$}\\
 & Improved action & Wilson action \\ \hline
1.0  &1.11(2)&1.32(1)   \\
1.35 &1.4(1)&1.5(2)    \\
1.55 &1.7(1)&2.1(1.1)      \\
1.70 &1.9(2)&2.2(1.5)    \\
1.90 &2.0(3) &2.2(1.5)    \\
2.0  &2.0(5) &2.4(1.6)    \\
\end{tabular}
\end{ruledtabular}
\end{table}  

\begin{table}[!h]
\caption{
\label{tab:results}
PIMC results for the string tension, $K$, symmetric and antisymmetric 
glueball masses $M_{S}$, $M_{A}$ and the mass ratio  $R_{M}$.}
\begin{ruledtabular}
\begin{tabular}{cccccc}
$\beta$& $a_{eff}$ &$K$ & $M_{S}/K$ & $M_{A}/K$ & $R_{M}$ \\ \hline
1.0  &0.3724 &0.329(2)& 5.10(5)   & 4.58(4)   & 1.11(2) \\ 
1.15 &0.2481 &0.189(3)& 7.8(1)    & 6.1(1)    & 1.28(3)  \\
1.25 &0.1884 &0.141(2)& 8.9(2)    & 6.4(1)    & 1.38(4)  \\
1.35 &0.1425 &0.112(5)& 10.8(5)   & 7.3(4)    & 1.4(1)  \\
1.40 &0.1239 &0.096(5)& 10.4(5)   & 6.7(3)    & 1.6(1)   \\
1.45 &0.1076 &0.067(4)& 13.9(9)   & 8.0(5)    & 1.7(2)   \\    
1.55 &0.0810 &0.045(3)& 16.6(1.0) & 9.8(6)    & 1.7(1)   \\    
1.60 &0.0702 &0.041(2)& 18.1(1.1) & 10.1(6)   & 1.8(1)   \\
1.65 &0.0608 &0.034(2)& 19.3(1.3) & 10.4(9)   & 1.8(2)   \\
1.70 &0.0527 &0.028(2)& 24.1(2.1) & 12.6(1.0) & 1.9(2)   \\ 
1.75 &0.0425 &0.023(2)& 22.0(2.2) & 11.6(1.1) & 1.9(3)   \\
1.80 &0.0395 &0.021(2)& 21.6(2.7) & 11.1(1.3) & 1.9(3)   \\
1.85 &0.0341 &0.016(2)& 23.4(2.5) & 12.3(1.3) & 1.9(3)   \\
1.90 &0.0295 &0.013(2)& 25.9(3.2) & 13.0(1.6) & 1.6(3)   \\
2.0  &0.0220 &0.010(2)& 26.0(4.5) & 12.9(2.2) & 2.0(5)   \\ 
2.10 &0.0164 &0.009(2)& 23.4(5.0) & 11.5(2.5) & 2.0(6)   \\   
2.20 &0.0122 &0.008(2)& 18.9(8.7) & 9.6(2.4)  &   \\
\end{tabular}
\end{ruledtabular}
\end{table}

\section{Summary and conclusions}
\label{sec:con}
In this work, we have applied the standard Euclidean path integral Monte 
Carlo method to obtain  results in the Hamiltonian limit of the 
Symanzik improved $U(1)_{2+1}$ lattice gauge theory on anisotropic lattices.
Monte Carlo results were obtained for the static quark potential, 
renormalized anisotropy, the string tension, and the lowest-lying 
glueball masses. The inter-quark potential with the improved action exhibits 
good rotational symmetry. We found that both the improved and standard Wilson 
action do 
equally well in extracting the static potential at large separations.
The improved discretization  allows substantially more accurate estimates 
of the string tension and the glueball masses.
The extrapolations to the Hamiltonian limit were performed by simple 
quadratic fits. 

In this limit the string tension displays an asymptotic 
behaviour which is in excellent agreement with the behaviour predicted by 
the  weak-coupling approximation of G{\" o}pfert and Mack \cite{gof82}. 
The scaling coefficient of the scaling curve 
for the mass gap was estimated  as roughly twice 
larger in magnitude than the weak-coupling prediction. This is an improvement 
over our previous unimproved results which were estimated larger by a  factor 
of 5-6. We believe that these estimates can be further improved by taking the 
renormalization of the couplings into account, i.e, computing the action 
beyond the tree-level.   

The weak-coupling behaviour of the antisymmetric mass gap was found 
to agree with the theoretical expectations. The parameters agree 
quite well with the results obtained from  previous numerical and analytic 
calculations. The mass ratio $R_{M}=M_{S}/M_{A}$ was observed to scale to
$2.007\pm 0.01$ in the continuum limit. No sign of a glueball bound state or a
sharp crossover between the levels was seen. This is in excellent agreement 
with the statement of G{\" o}pfert and Mack \cite{gof82}, that the 
continuum limit corresponds to a theory of free bosons, where $R_{M} =2$ 
exactly. It also shows very clearly the value of using an improved action, in
giving more rapid convergence to the continuum limit, as well as improved 
accuracy.

Taking  advantage of the improved discretization on anisotropic lattices, we 
aimed to apply the Euclidean path integral Monte Carlo approach to examine the 
Hamiltonian limit of $U(1)$ theory. The results obtained here clearly 
demonstrate that PIMC is a more reliable technique than other quantum Monte 
Carlo methods, such as Green's Function Monte Carlo, which gave only 
qualitative estimates of the string tension and the mass gap.

Although suffering from the disadvantage 
that it requires an  extrapolation to the 
anisotropic limit, PIMC is the preferred Monte Carlo technique 
for obtaining reliable results in the Hamiltonian formulation, just as 
in the Euclidean formulation. In order to make the PIMC method a  more  
valuable tool in Hamiltonian lattice gauge theories, it will be crucial to 
show that it allows one to treat eventually matter fields as well as gauge 
fields especially in the non-Abelian case.  

\acknowledgments

This work was supported by the Australian 
Research Council. We are  grateful for access to the
computing facilities of the Australian Centre for Advanced Computing and
Communications (ac3) and the Australian Partnership for Advanced
Computing (APAC).

\end{document}